\def\year{2022}\relax
\documentclass[letterpaper]{article} % DO NOT CHANGE THIS
\usepackage{aaai22}  % DO NOT CHANGE THIS
\usepackage{times}  % DO NOT CHANGE THIS
\usepackage{helvet}  % DO NOT CHANGE THIS
\usepackage{courier}  % DO NOT CHANGE THIS
\usepackage[hyphens]{url}  % DO NOT CHANGE THIS
\usepackage{graphicx} % DO NOT CHANGE THIS
\usepackage{natbib}  % DO NOT CHANGE THIS AND DO NOT ADD ANY OPTIONS TO IT
\usepackage{caption} % DO NOT CHANGE THIS AND DO NOT ADD ANY OPTIONS TO IT
\DeclareCaptionStyle{ruled}{labelfont=normalfont,labelsep=colon,strut=off} % DO NOT CHANGE THIS
\usepackage[ruled,vlined]{algorithm2e}
\usepackage{subcaption}

\newtheorem{definition}{Definition}[section]
\usepackage{amsmath,amssymb,amsfonts}
\usepackage{subcaption}
\usepackage{multirow}

\usepackage{array}
\usepackage{booktabs}
\usepackage{graphicx}
\usepackage{threeparttable}
\usepackage{wasysym}

\usepackage[framemethod=tikz]{mdframed}

\makeatletter
\def\ProblemSpecBox{
  \@ifnextchar[\ProblemSpecBox@opt{\ProblemSpecBox@noopt}}
\def\ProblemSpecBox@opt[#1]#2{
  \protected@edef\@currentlabelname{#1}
  \protected@edef\@currentlabel{#1}
  \begin{mdframed}[
    innerlinewidth=0.5pt,
    innerleftmargin=10pt,
    innerrightmargin=10pt,
    innertopmargin = 10pt,
    innerbottommargin=10pt,
    skipabove=\dimexpr\topsep+\ht\strutbox\relax,
    roundcorner=5pt,
    frametitle={#2},
    frametitlerule=true,
    frametitlerulewidth=1pt]
}
\def\ProblemSpecBox@noopt#1{
  \ProblemSpecBox@opt[#1]{#1}
}
\def\endProblemSpecBox{
  \end{mdframed}
}
\makeatother

\title{\textsc{\mdseries Beas}: Blockchain Enabled Asynchronous \& Secure Federated Machine Learning}
\author {
    Arup Mondal,
    Harpreet Virk,
    Debayan Gupta
}
\affiliations {
    Ashoka University \\
    arup.mondal\_phd19@ashoka.edu.in, harpreet.virk@alumni.ashoka.edu.in, debayan.gupta@ashoka.edu.in
}

\usepackage{bibentry}
\begin{document}

\maketitle

\begin{abstract}
Federated Learning (FL) enables multiple parties to distributively train a ML model without revealing their private datasets. However, it assumes trust in the centralized \textit{aggregator} which stores and aggregates model updates. This makes it prone to gradient tampering and privacy leakage by a malicious aggregator. Malicious parties can also introduce backdoors into the joint model by poisoning the training data or model gradients. To address these issues, we present \textsc{Beas}, the first blockchain-based framework for $N$-party FL that provides strict privacy guarantees of training data using gradient pruning (showing improved differential privacy compared to existing noise and clipping based techniques). Anomaly detection protocols are used to minimize the risk of data-poisoning attacks, along with gradient pruning that is further used to limit the efficacy of \textit{model-poisoning} attacks. We also define a novel protocol to prevent premature convergence in heterogeneous learning environments.% , and also some resistance to \textit{data-poisoning} and \textit{model-poisoning} attacks. Anomaly detection protocols are used to minimize the risk of data-poisoning attacks. To preserve the confidentiality of the training data, we employ gradient pruning (showing improved differential privacy compared to existing noise and clipping based techniques).  
We perform extensive experiments on multiple datasets with promising results: \textsc{Beas} successfully prevents privacy leakage from dataset reconstruction attacks, and minimizes the efficacy of poisoning attacks. Moreover, it achieves an accuracy similar to centralized frameworks, and its communication and computation overheads scale linearly with the number of participants. 
% We also discuss practical use cases that use \textsc{Beas} as a general FL framework, % -- federated transfer learning, 
% (ii) synthetic data aggregation, (iii) federated one-shot dataset distillation, and (iv) privacy preserving machine learning, 
% demonstrating wide applicable potential of the proposed framework.

\end{abstract}

%
%% ---- Section ----
%
\section{Introduction}\label{sec:intro}

Robust Machine Learning (ML) models require large amount of heterogeneous training data to obtain accurate results of any practical significance. In most scenarios, this data is often scattered across mutually distrusting entities that cannot directly share their secret private data with each other, or with a centralized aggregator, due to privacy regulations that restrict centralized collection~\cite{sav2020poseidon,ramachandran2021s++}. Federated Learning (FL)~\cite{mcmahan} enables multiple parties to distributively learn a shared model without revealing their private data; each party trains a \textit{local} model using their own data, and exchanges only the model gradients with a centralized FL \textit{server} or \textit{aggregator}. The \textit{aggregator} is responsible for storage and exchange of these gradients. It also periodically merges them, generally by taking their average, to generate a new \textit{global} model that is then used in subsequent \textit{local} rounds for pre-training. The aggregator is thus a central player that potentially represents a single point of failure~\cite{sav2020poseidon}. 

~\cite{bonawitz} proposed an improved FL framework by designing a hierarchical network of FL aggregators, each of which controls its own sub-population headed by a centralized coordinator. Although this reduces the risk of a malicious aggregator by distributing the task across multiple sub-networks, the architecture is not inherently free from centralized dependency~\cite{sav2020poseidon}. 

~\cite{song} showed that model gradients can sometimes unintentionally memorize features from training data, which can be exploited by a malicious aggregator to reconstruct sensitive information about it. This can be solved using various cryptographic techniques~\cite{perry2014systematizing,he,di2014practical,gupta2016using,mood2016frigate}, though these significantly impact the efficiency of the framework~\cite{phong}. Differential privacy (DP)~\cite{dwork} techniques like addition of noise or clipping of gradients can also be used to limit leakage of privacy. However, obfuscating model gradients using DP techniques often impacts accuracy~\cite{lyu,abadi}. Moreover, the aggregator cannot examine the data used for training: malicious parties can train their \textit{local} models using poisoned datasets, or tamper with their gradients, to introduce backdoors into the shared model~\cite{jagielski,bagdasaryan}. A malicious aggregator can also skew the shared model by biasing contributions of preferred parties~\cite{li}. Data poisoning attacks have been addressed with anomaly detection algorithms such as \textit{l-nearest aggregation}~\cite{chen} and Multi-KRUM~\cite{blanchard}, where deviation or performance of updates is evaluated relative to the majority. However, ~\cite{bagdasaryan} demonstrates attacks where an adversary can use \textit{constrain-and-scale} techniques to generate gradients that can evade anomaly detection. 

Researchers have investigated eliminating the centralization in FL using decentralized frameworks. However, most of the current work primarily addresses only a subset from - secure design, byzantine tolerant gradient aggregation, protection from poisoning attacks, data privacy, client motivation, fairness guarantees, scalability, and network efficiency. Currently, only~\cite{chen,shayan} address a majority of these issues, but not all. Moreover, prior work does not consider the improved privacy protection offered by gradient pruning~\cite{zhu}, or the critical issue of premature convergence: the \textit{global} model can forget previously learned features if the \textit{local} models prematurely converge while training on a large dataset.

We propose \textsc{Beas}, the first blockchain enabled framework that allows collaborative training and evaluation of ML tasks in a distributed setting with strict privacy guarantees and security against adversarial clients. Our primary motivation to use blockchain are smart contracts, which can function as a trusted distributed application and be used for storage, exchange, and merging of gradients, eliminating centralized dependence entirely. Using a fully-decentralized blockchain, \textsc{Beas} can enable secure FL executions with different types of layered architectures, such as feed-forward NN and CNN, on datasets that are heterogeneously distributed among $N$ parties.
% \textsc{Beas} uses score based gradient anomaly detection to protect the \textit{global} model from data poisoning attacks. Gradient pruning is used to protect the shared model from model poisoning attacks with minimal impact on model accuracy. To protect client data privacy, gradient pruning is further used for improved differential privacy compared to existing noise-addition based techniques used in prior work; gradient pruning helps minimize the reconstruct-ability of training data from shared model gradients under a Deep Leakage from Gradient attack~\cite{bagdasaryan}. 
Moreover, \textsc{Beas} is designed to be scalable with parallel FL models running on the same platform using multi-channel blockchain architecture. This can be useful in various research and production applications. %We also examine research on statistical heterogeneity for solutions to minimize the effect of non-i.i.d. dataset distribution, and use federated averaging to aggregate and merge client gradient contributions.

% \noindent
\paragraph{\textbf{Contributions.}}
% We summarize 
Our main contributions as follows:
\begin{itemize}
    \item Blockchain-based framework for decentralized $N$-party (unbounded $N$) FL ensuring strict privacy guarantees of training data using gradient pruning based DP, and resistance to poisoning attacks.
    
    \item Minimize risk of data poisoning using a combination of protocols to identify adversaries: (i) Multi-KRUM ~\cite{blanchard} is used to guarantee resiliency from independent adversaries; and (ii) FoolsGold~\cite{fg} is used to identify Sybil groups.
    
    \item Implement and compare various DP techniques~\cite{dwork} to prevent direct leakage of training data from shared gradients. Our experiments show gradient pruning (GP) is more effective than existing DP techniques: it prevents reconstruction of training data from shared model gradients with minimal impact on performance, and defends against model poisoning. GP has not been used in prior work for privacy (GP's primary use: gradient compression). We also evaluate the efficiency of GP against model-poisoning attacks. To the best of our knowledge, \textsc{Beas} is the first approach for decentralized privacy-preserving FL to use GP for improved differential privacy, and improved defense against model poisoning.
    
    \item \textsc{Beas} is first to explore premature convergence in decentralized FL; we train using small clusters of data to prevent this.  We're also first to support parallel FL collaboration on multiple tasks using the same platform with multi-channel blockchain architecture.
    
    \item We perform extensive experiments on multiple datasets and different NN architectures with promising results: \textsc{Beas} achieves training accuracy on par with both -- centralized and non-privacy preserving decentralized approaches. \textsc{Beas} can train a 3-layer NN with 64 neurons per hidden-layer with the training dataset split amongst 20 participating clients on the MNIST~\cite{lecun-mnisthandwrittendigit-2010} dataset in 8.73 minutes, and on the Malaria Cell Image ~\cite{malaria} dataset in 16.11 minutes. 
\end{itemize}

For ease of access, all of our code and experiments are available at: \textcolor{blue}{\url{https://github.com/harpreetvirkk/BEAS}}.

% \paragraph{\textbf{Organization of the Paper:}} 
% \S\ref{sec:backprelim} describes technical preliminaries and backgrounds. 
% In \S\ref{sec:beas}, we describe the problem statement, followed by the proposed \textsc{Beas} framework. 
% The experimental evaluation, as well as the security and privacy analysis, are presented in \S\ref{sec:experiment} and \S\ref{sec:secpriana} respectively. 
% % Section~\ref{sec:companalysis} shows a detailed comparison between \textsc{Beas} and existing state-of-the-art frameworks. 
% % We examine \textsc{Beas} as a general privacy-preserving framework in section~\ref{sec:app}. 
% %Section~\ref{sec:relatedwork} provides a brief overview of related works, and 
% We conclude in \S\ref{sec:conclusion}.

\subsection{Comparative Analysis}\label{sec:companalysis}

We compare \textsc{Beas} against existing \textit{state-of-the-art} frameworks for decentralised FL. \textsc{Beas} uses multi-channel permissioned blockchain to store all model gradients, which enables rapid scalability, auditability, transparency, and trust amongst collaborating entities. Differential privacy~\cite{dwork} is used to obfuscate model gradients to prevent leakage of sensitive information of private data, and FoolsGold~\cite{fg} with Multi-KRUM~\cite{blanchard} is used to provide resiliency from adversarial data poisoning attacks~\cite{jagielski}. Gradient pruning is also used to limit the efficacy of model poisoning attacks~\cite{bagdasaryan}, where malicious clients use \textit{constrain-and-scale} techniques to evade gradient anomaly detection algorithms such as Multi-KRUM. Table~\ref{table:companalysis} shows the comparative analysis of the proposed \textsc{Beas} framework against other
existing frameworks.

\begin{table*}[ht]
\centering
\small
\caption{Comparison between various privacy-preserving federated learning framework.}
\label{table:companalysis}
\newcommand*\rot[1]{\hbox to1em{\hss\rotatebox[origin=br]{-60}{#1}}}
\newcommand*\feature[1]{\ifcase#1 $\square$\or$\boxtimes$\or$\blacksquare$\or\Circle\or\LEFTcircle\or\CIRCLE\or--\or$\times$\or$\checkmark$\or \fi}
\newcommand*\g[2]{\feature#1&\feature#2}
\newcommand*\f[3]{\feature#1&\feature#2&\feature#3}
\newcommand*\h[9]{\feature#1&\feature#2&\feature#3&\feature#4&\feature#5&\feature#6&\feature#7&\feature#8&\feature#9}
\newcommand*\e[6]{\feature#1&\feature#2&\feature#3&\feature#4&\feature#5&\feature#6}
\makeatletter
\newcommand*\ex[7]{#1\tnote{#2}&#3&\g#4&\g#5&\f#6&\e#7&\expandafter\h\@firstofone
}
\makeatother
\newcolumntype{G}{c@{}c}
\newcolumntype{F}{c@{}c@{}c}
\newcolumntype{H}{c@{}c@{}c@{}c@{}c@{}c@{}c@{}c@{}c}
\newcolumntype{E}{c@{}c@{}c@{}c@{}c@{}c}
\begin{threeparttable}
\begin{tabular}{@{}lc G  G F E H@{}}
\toprule
Framework  & Comms$^{\dag}$ & \multicolumn{2}{c}{\shortstack{Threat \\ Model}} & \multicolumn{2}{c}{\shortstack{Privacy \\ Guarantees}} & \multicolumn{3}{c}{\shortstack{Security \\ Guarantees}} & \multicolumn{6}{c}{\shortstack{Techniques Used}}  & \multicolumn{9}{c}{\shortstack{Features and \\ Code Availability}} \\ %& \multicolumn{6}{c}{\shortstack{Experimental \\ Dataset}} \\
\midrule

% rotated items
&& \rot{Participants}
 & \rot{Aggregator}
 & \rot{Inference}
 & \rot{Training}
 & \rot{Data Poisoning}
 & \rot{Model Poisoning}
 & \rot{Byzantine Attack}
 & \rot{HE}
 & \rot{TP}
 & \rot{SS+AE}
 & \rot{FE}
 & \rot{DP}
 & \rot{Blockchain}
 & \rot{Identity Privacy}
 & \rot{Scalability}
 & \rot{Statistical Heterogenity} 
 & \rot{Asynchronous Updates}
 & \rot{Dynamic Participants}
 & \rot{Decentralized} 
 & \rot{Premature Convergence}
 & \rot{Reward Mechanism} 
 & \rot{Open-Source} \\
\midrule

\ex{BinDaaS~\cite{bhattacharya}}{} {3 rounds} {06} {33} {333} {333335} {535333337} \\ %{333333}\\

\midrule

\ex{PiRATE~\cite{zhou}}{} {3 rounds} {16} {44} {533} {333335} {553333357} \\ %{333333}\\

\midrule

\ex{BAFFLE~\cite{ramanan}}{} {3 rounds} {26} {55} {533} {333335} {553333357} \\ %{333333}\\

\midrule

\ex{Li et al.~\cite{li}}{} {3 rounds} {16} {33} {533} {333335} {535333357} \\ %{533333}\\

\midrule

\ex{LearningChain~\cite{chen}}{} {3 rounds} {26} {55} {444} {333355} {553333357} \\ %{533533}\\

\midrule

\ex{Biscotti~\cite{shayan}}{} {3 rounds} {26} {55} {544} {333355} {555333557} \\ %{533353}\\

\midrule

\ex{Truex et al.~\cite{truex2019hybrid}}{} {3 rounds} {21} {33} {333} {353353} {343333337} \\ %{533333}\\

\midrule

\ex{Bonawitz et al.~\cite{bonawitz}}{} {3 rounds} {21} {33} {333} {335333} {533343337} \\ %{333333}\\

\midrule

\ex{PySyft~\cite{ryffel2018generic}}{} {2 rounds} {01} {33} {333} {533333} {333333337} \\ %{333335}\\

\midrule

\ex{FLTrust~\cite{cao2020fltrust}}{} {2 rounds} {21} {45} {555} {333333} {553533337} \\ %{53333}\\

\midrule

\ex{POSEIDON~\cite{sav2020poseidon}}{} {2 rounds} {21} {55} {333} {533333} {553335337} \\ %{53333}\\

\midrule

\ex{Shokri et al.~\cite{shokri2015privacy}}{} {1 round} {00} {33} {333} {333333} {453343337} \\  %{53333} \\

\midrule

\ex{PATE~\cite{papernot2018scalable}}{} {1 round} {00} {33} {333} {333333} {343333337} \\ %{53333} \\

\midrule

\ex{HybridAlpha~\cite{DBLP:conf/ccs/XuBZAL19}}{} {1 round} {21} {54} {433} {333553} {553353337} \\ %{533333}\\

\midrule

\ex{\textbf{\textsc{Beas}(This Work)}}{} {1 round} {26} {55} {555} {333355} {555555558} \\ %{555533}\\

\bottomrule
\end{tabular}

\begin{tablenotes}
\item 
\textsc{Beas} proposes efficient privacy-preserving federated learning framework in decentralized settings with stronger security and privacy guarantees. 
\textsc{Beas} also provides an extensive evaluation and comparison (1) over the large number of datasets, (2) provides an extensively compares with related work, (3)  provides newer insights for future directions and a number of interesting application. 

\item ``HE`` is homomorphic encryption'; ``TP'' is Threshold-Paillier system; ``SS+AE'' secret sharing with key agreement protocol and authenticated encryption scheme; ``FE'' is functional encryption; and ``DP'' is Differential Privacy. 
\item $\dag$ includes the number of communication rounds required in one epoch at the training phase between the aggregator and the participant.
\item \feature0 \text{denotes honest party}; \feature1 \text{denotes semi-honest party}; \text{\feature2} \text{denotes dishonest party};  $\text{\feature3} \; \text{denotes does not provides property}$;  $\text{\feature4} \; \text{denotes partially provides property}$;  $\text{\feature5} \; \text{denotes provides property}$.
\end{tablenotes}

\end{threeparttable}
\end{table*}
\section{Related Work}\label{sec:relatedwork}

Federated learning (FL)~\cite{mcmahan} has emerged as a promising approach to collaboratively train a model by exchanging model parameters with a central aggregator, instead of the actual training data. However, parameter exchange may still leak a significant amount of private data. To overcome this leakage problem, several approaches have been proposed based on differential privacy~\cite{lyu}, multi-party computation~\cite{mondal2022scotch,DBLP:conf/ccs/XuBZAL19,bonawitz,ryffel2018generic}, fully homomorphic encryption~\cite{truex2019hybrid}, Trusted Execution Environment~\cite{mondal2021poster,mondal2021flatee}, etc. However, due to the extensive use of cryptographic operations, these protocols remain too slow for practical use. Furthermore, in those settings, the aggregator is a central player, which also potentially represents a single point of failure~\cite{sav2020poseidon}. 

To overcome this single point of failure, Shae et al.~\cite{shae} considers a theoretical implementation of distributed learning and transfer learning with blockchain smart contracts to design a distributed parallel computing architecture, and~\cite{lu} proposes crowd-sourcing of ML tasks on public blockchains where nodes are incentivized for contributing their computational resources. Similarly,~\cite{harris} uses blockchain to build a public dataset, and smart contracts are used to host a continuously updated model. However, their focus is not on sensitive datasets: data needs to be published on the blockchain~\cite{lu} or is shared freely amongst nodes for efficient computation~\cite{harris}.

A number of approaches~\cite{lu,yong,bhattacharya,ide,majeed,wang,kuo} try to eliminate the centralized dependence in a learning framework using blockchain, these are still subject to poisoning attacks by malicious nodes that train their local models using poisoned datasets~\cite{jagielski} or tamper with their model gradients~\cite{bagdasaryan}. \cite{li} presents a committee consensus mechanism to validate gradients where a leader is selected based upon performance in the previous round. However, this can result in poor performance of an honest node when the distribution of the leader’s dataset is more similar to that of the malicious node than the honest node. ~\cite{song} further demonstrates membership inference attacks that can reveal whether a specific data point was used to train a given model, and how the models themselves can unintentionally memorize training inputs.

The current state-of-the-art for decentralized FL frameworks~\cite{shayan,chen} tries to eliminate the poisoning attacks, inference attacks, and membership attacks by using the several cryptographic techniques. However, due to the extensive use of cryptographic schemes, these frameworks remain too slow for practical use. The main advantage \textsc{Beas} has over these approaches is the communication efficiency, and stronger data privacy and security guarantees (see Table~\ref{table:companalysis}). The existing approaches need more than one round of communication in the local model aggregating phase, \textsc{Beas} only incurs a single round. Hence \textsc{Beas} can be used to train machine learning models faster as demonstrated in the experimental section.

\section{Background and Preliminaries}\label{sec:backprelim}

% \paragraph{\textbf{Federated Learning.}}
\subsection{Federated Learning}
Federated learning (FL)~\cite{mcmahan} is a distributed machine learning approach that enables collaborative model training on decentralized data. %FL is suitable for use cases that involve sensitive datasets such as healthcare, financial services, etc. 
The training setup comprises multiple clients with private datasets that want to collaboratively train a shared global model, but do not wish to expose their secret private dataset to other participating clients~\cite{Yang2019}. To do so, every client trains a model locally with their own data, and exchanges only the model parameters with an FL \textit{aggregator} instead of directly sharing the private training data. The aggregator merges the received local gradients (generally by averaging) to generate the new global model, which is then sent back to all the participants to use as a pre-training model for the next iteration of local training. The process repeats until desired performance is achieved, or \textit{ad infinitum}.

%The FL training process is carried out as follows:

%\begin{enumerate}
%    \item Participants $\mathcal{F}_{1}, \mathcal{F}_{2}...\mathcal{F}_{N}$ compute their gradients with respect to local datasets  $\mathcal{D}_{1}, \mathcal{D}_{2}...\mathcal{D}_{N}$, and might employ different techniques such as secret sharing, homomorphic encryption, and differential privacy to conceal results before sharing it to an honest but curious server $\mathcal{S}$. 
%    \item The server $\mathcal{S}$ then partakes in a secure aggregation of the received updates, without learning anything about the underlying updates.
%    \item The server $\mathcal{S}$ sends the aggregated update to each of the participants $\mathcal{F}_{i}$  $\forall 1 \le i \le N$.
%    \item After receiving the updates from the server $\mathcal{S}$ each participant $\mathcal{F}_{1}, \mathcal{F}_{2}...\mathcal{F}_{N}$ updates their local model after decrypting the aggregated update. 
%\end{enumerate}

\subsection{Blockchain}
% \paragraph{\textbf{Blockchain.}}
Blockchain is a decentralized immutable ledger of records, called blocks, that are sequentially linked using cryptography and maintained across a network of presumably distrustful peers \cite{narayanan,nakamoto}. Every peer maintains a copy of the ledger, eliminating the dependency on a centralized authority. Blockchains typically allow execution of programmable scripts. Crypto currencies such as Bitcoin use these scripts to validate transactions~\cite{nakamoto}. HyperLedger Fabric~\cite{androulaki} extends this capability with the concept of smart contracts that can function as a trusted decentralized application (DApp)~\cite{androulaki}. In the context of decentralized FL, smart contracts can be used for storage, exchange, and merging of model gradients, replacing the dependence on a centralized aggregator. We implement \textsc{Beas} using HyperLedger Fabric~\cite{androulaki}, which is an open source enterprise-grade permissioned blockchain framework, for the following reasons: (a) permissioned blockchain architecture, (b) easily scalable, (c) multi-channel blockchain design, (d) transaction level consensus, (e) no crypto-currency requirement, and (f) no proof-of-work/stake.

HyperLedger  Fabric~\cite{androulaki} is  an  open  source  enterprise-grade  permissioned  blockchain  framework. At a high level, it is comprised of the following modular components~\cite{hyp-ibm2,androulaki}: 

\begin{enumerate}
    \item Anything that can have value, state, and ownership are called \textit{Assets}, and are represented as a set of key-value pairs. 
    \item \textit{Ledger} records the state and  ownership of an asset at a given point in time, as well as a log history that records every transaction.
    \item \textit{Smart contracts (or chaincode)} is software that comprises the business logic of the framework; that is, it defines the assets and the applicable transaction functions.
    \item \textit{Endorsing Peers} are a fundamental component of the entire framework. They host and access the ledgers, endorse transactions, interface with applications, and execute smart contracts within a secure container environment (e.g. Docker) for isolation. 
    \item \textit{Endorsement policy} defines the necessary and sufficient conditions for a valid transaction endorsement.
    \item \textit{Channels} are logical structures formed by a collection of participating peers, allowing a group of peers to create a separate ledger of transactions.
    \item \textit{Membership service provider (MSP)} is responsible for associating entities in the network with cryptographically generated identities.
    \item \textit{Ordering service} packages transactions into blocks, and establishes consensus on the order of transactions. It guarantees the delivery of transactions in the network using a peer-to-peer gossip protocol.
\end{enumerate}

We select Fabric over other blockchain platforms for \textsc{Beas} due to the following reasons \cite{hyp-ibm,androulaki}: 

\textit{Permissioned blockchain: } Anyone can join public permission-less blockchains networks, including anonymous and pseudonymous users. Fabric's permissioned blockchain design requires all participants to have cryptographically generated anonymous identities, which is useful in our case as malicious parties can be removed from the network when their adversarial intent is detected. 

\textit{Scalability: }Most public blockchains require all the nodes on the network to process transactions, which results in low transaction throughput and impacts scalability. In Fabric, network designers can plug in the requisite nodes as per their immediate requirements, making the framework highly scalable. 

\textit{Multi-Channel: }The need for data partitioning on the blockchain is essential for most business use-cases due to competitiveness, protection laws, and regulation on confidentiality of personal data. In Fabric, Channels allow for data to go to only the parties that need to know. In our case, we use multi-channel blockchain to facilitate parallel collaboration with multiple models on the same platform by setting up each channel with its own ML task, and a separate blockchain ledger to store the shared model gradient blocks.

\textit{Transaction Consensus: }While most blockchain frameworks require consensus on a ledger state before transaction approval, Fabric relies on a transaction level consensus; in other words, the entire block doesn't require validation before being approved, but only the transaction.

\textit{No Crypo-Currency required: } Public blockchain frameworks like Ethereum \cite{wood} require crypto-currencies to execute chaincode functionality, which is redundant in our use case. Fabric can be used without a cryptocurrency.

\textit{No Proof of work/stake: } Fabric does not depend upon miners working day and night to solve problems for block validation, which requires lots of computing power and is energy-intensive.

\subsection{Differential Privacy}\label{subsec:appendp}
% Zhu et al.~\cite{zhu} demonstrates that it is possible to reconstruct sensitive information from generated model gradients. 
Differential privacy~\cite{dwork} can be used to ensure that the presence (or absence) of any given element in a dataset does not result in the generation of vastly different model gradients (which could potentially be used to reveal private information from shared model gradients in FL). Differential privacy in the context of privacy-preserving FL, is defined in Definition~\ref{def:1}. 
% Existing work majorly relies on output perturbation, by adding noise or clipping the norm of the gradients, for differentially private sharing of model gradients~\cite{lyu,shayan,abadi,balle}.

\begin{definition}[Differential Privacy~\cite{dwork}]
{
\label{def:1}
A randomized function $K$ gives  $\epsilon-$differential privacy if for all model gradients $G_1$ and $G_2$, generated by training on datasets $D_1$ and $D_2$ differing on at most one element, the probability of function $K$ resulting in an output $S$ on $G_1$ is close to the probability of function $K$ resulting in the same output $S$ on $G_2$ as follows: 
$$Pr[K(G_1) \in S ] \leq exp(\epsilon) \times Pr[K(G_2) \in S ]$$ where all $S \subseteq Range(K)$.

% \begin{equation}
%     Pr[K(G_1) \in S ] \leq exp(\epsilon) \times Pr[K(G_2) \in S ] 
% \end{equation}

% where all $S \subseteq Range(K)$.
}
\end{definition}

Existing work majorly relies on output perturbation, by adding noise or clipping the norm of the gradients, for differentially private sharing of model gradients~\cite{lyu,shayan,abadi,balle}:

% \noindent
\textit{Addition of Noise:} When the query is a function $f$, and the database is $X$, the true answer is the value $f(X)$. The mechanism $K$ adds appropriately chosen random noise to the true answer to produce what we call the response~\cite{dwork}. Generally, a Gaussian or Laplace distribution is used for generating the noise for gradient perturbation.

% \noindent
\textit{Gradient Clipping:} Gradient value clipping involves clipping the derivatives of the loss function to have a given value if a gradient value is less than a negative threshold, or more than the positive threshold~\cite{brownlee}.

% \noindent
\textit{Gradient Pruning:}  In gradient pruning, gradients with small magnitudes are pruned to zero. The sparsity in the gradients increases as more and more gradient values are pruned, making attacks on data privacy hard~\cite{tsuzuku,lin}.

\section{\textsc{\mdseries Beas} Overview}\label{sec:beas}

\textsc{Beas} aims to achieve secure and efficient $N$-party machine learning while ensuring strict privacy guarantees that prevent leakage of sensitive information from shared model gradients, as well as ensure resiliency from adversaries. We consider a setting where $N$ mutually distrustful clients $\mathcal{C}_{1}, \mathcal{C}_{2}$... $ \mathcal{C}_{N}$, that hold private datasets $\mathcal{D}_{1}, \mathcal{D}_{2}$... $\mathcal{D}_{N}$ respectively, want to collectively train a shared \textit{global} model $\mathcal{M_{G}}$ without exposing their secret private dataset $\mathcal{D}_{i}$ to other participants. This need for collaboration arises because a model trained only on an individual’s data would exhibit poor performance, but a model trained by all participants will have near-optimal performance~\cite{shayan, Yang2019}. We also assume that the NN training parameters
are known to all clients: the model architecture, hyper-parameters,
optimization algorithm, and the learning task of the system (these can be requested from the endorsing peers). We target horizontal learning, where every user contributes data containing the same feature space, \textit{i.e.} data with same columns, but different/overlapping rows.

\subsection{Threat Model}

We assume \textit{curious and colluding} clients, who may collude to try to acquire sensitive information from other clients' private training datasets by inspecting the model gradients that are publicly published on the shared ledger. Moreover, clients may be adversarial, and contribute poisoned updates to introduce backdoors into the shared model. However, colluding clients cannot gain influence without acquiring sufficient stake. The adversary may control more than one client, as in Sybil attacks~\cite{douceur}, with the intent to take control of the blockchain consensus by acquiring a majority stake (by creating false identities). However, we assume that they can not control more than $f$ clients out of $N$ total clients, when $2f+2 <N$. Moreover, although adversaries may be able to increase the number of clients under their control, they cannot artificially increase their stake in the system except by either providing valid updates that pass Multi-KRUM~\cite{blanchard}, or by modifying their gradient updates to evade anomaly detection as in a model poisoning attack~\cite{bagdasaryan}. We further assume that the intent of the adversary is to harm the performance, or introduce backdoors, into the shared \textit{global} model, or leak private information about the used training dataset. For the purpose of this work, we limit the adversaries to label flipping attacks~\cite{labelflip}, pixel-pattern backdoor attacks~\cite{bagdasaryan}, and deep leakage from gradients using reconstruction attacks~\cite{zhu}.

\textsc{Beas}'s operating steps are describes in Scheme~\ref{alg:beasframe}. Algorithm~\ref{alg:beas} describes the high level required steps and operations of our proposed \textsc{Beas} framework for the decentralized federated machine learning training in $N$-party settings.

\begin{ProblemSpecBox}[1]{Scheme 1: \textsc{\mdseries Beas} Framework\label{alg:beasframe}}
% \scriptsize

    \textbf{Input:}  Client $\mathcal{C}_i$ for $i \in \{1,\cdots,K\}$ holds its private dataset $\mathcal{D}_i$. \\
    \textbf{Output:} Client $\mathcal{C}_i$ for $i \in  \{1,\cdots,K\}$ obtain the collaboratively learned \textit{global} model, $\mathbb{M}_{G}$. 
    \begin{enumerate}
        \item Clients create cryptographically anonymous identities using the MSP.
        
        \item To begin training process, any client $\mathcal{C}_i\in \mathcal{C}_K$ can set up a new channel $c$, define the training parameters and the model architecture, and generate a genesis block $\mathcal{M}_g$ by training on their own private data $D_i$ locally. $\mathcal{M}_g$ is uploaded as the first \textit{global} block on the channel ledger $\mathcal{L}_c$.
        
        \item Other clients connect to the EP to request the latest \textit{global} block from the channel ledger. They use the requested block to initialize a pre-trained model, and update it by training on their own private datasets $D_i$ to generate new \textit{local} gradients.
        
        \item Client $\mathcal{C}_i$ sends their \textit{local} gradients to the EP, which creates a new \textit{local} block and shares it with the Ordering service.
        
        \item The ordering service establishes consensus on the ordering of blocks, and consequently commits them onto each EP’s ledger using a peer-to-peer gossip protocol. 
        
        \item The EP checks if the number of queued \textit{local} blocks is $\geq$ than  the  merge  threshold.  If  true,  the Merge chaincode is triggered to evaluate the quality of each \textit{local} block by calculating the anomaly detection scores, and aggregating blocks using federated averaging to generate a new \textit{global} block, which is then sent to the channel ledger. 
        
        \item Steps 3 to 6 get repeated until desired accuracy for the shared \textit{global} is achieved or \textit{ad-infinitum}.
    \end{enumerate}

\end{ProblemSpecBox}

\begin{algorithm}[ht]
% \scriptsize
\KwIn{\textsc{Beas} with $K$ Local Blocks generated by $K$ clients, for $P_k$ the set of indexes of data points on client $k$, and $n_k = {\mid p_k \mid}$. $f_i(w) = l(x_i, y_i;w)$ is the loss of the prediction on example $(x_i, y_i)$ made with model parameters $w$. When $K>t$, where $t$ is the merging threshold, the merge chaincode is triggered to generate the Global Block. $\mathcal{M}_{FL} :=$ Machine learning algorithms to be trained; $\epsilon :=$ differential privacy guarantee; $\mathcal{C}_{1 \rightarrow N} :=$ Set of $N$ participants, where $\mathcal{C}_i$ holds its own dataset $\mathcal{D}_i$; $c :=$ local training cluster size.}
\KwOut{Trained Global Model $\mathcal{M_G}$}    
\caption{\textsc{\mdseries Beas} Training Algorithm}\label{alg:beas}
\DontPrintSemicolon
    
    % Set Function Names
    \SetKwFunction{FSetup}{\textbf{ChannelInitialize}}
    \SetKwFunction{FModelAggregate}{\textbf{ModelAggregate}}
    \SetKwFunction{FLocalTraining}{\textbf{LocalTraining}}
    
    \SetKwProg{Fn}{}{:}{}
    \Fn{\FSetup{$t$}}{
        \For{each training round}
        {
            $K = 0$ \;
            \For{each client $\mathcal{C}_k$ in parallel}
            {
                $\mathcal{M}_L^k$ = LocalTraining($ \mathcal{M}_G, k, w$) \;
                $K=K+1$ \;
                \If{$K>t$}
                {
                    $\mathcal{M}_G \rightarrow$  ModelAggregate($\mathcal{M}_L$, w)
                }
            }
        }
        % \KwRet \;
    }
    % \;
    
    \SetKwProg{Fn}{}{:}{}
    \Fn{\FLocalTraining{$ \mathcal{M}_G, k, w$}}{
        pretrain $\mathcal{M}_L$ using $ \mathcal{M}_G$ \;
        $\mathcal{M}_L^k \rightarrow \frac{1}{n_k} \sum_{i \in P_k} f_i(\mathcal{M}_L^k ) + \epsilon$ \tcp*{\scriptsize run on client k}
        \KwRet $\mathcal{M}_L^k$\;
    }
    % \;
    
    \SetKwProg{Fn}{}{:}{}
    \Fn{\FModelAggregate{$\mathcal{M}_L,  w$}}{
        % MKScore($F_k(w)^i$) =  \sum_{i\rightarrow j} \mid\mid $w_i$ - $w_j$ \mid\mid^2 \;
        MKScore($\mathcal{M}_L^i$) =  $\sum_{i\rightarrow j} \mid\mid \mathcal{M}_L^i - \mathcal{M}_L^{j} \mid\mid^2$ \;
        
        \For{top $m$ blocks by score}
        {
            $\mathcal{M}_G  \rightarrow \sum_{k=1}^{K} \frac{n_k}{n}F_k(\mathcal{M}_L^i)$
        }
        \KwRet $\mathcal{M}_G$\;
    }
\end{algorithm}

\subsection{System Architecture}

\begin{figure}[ht]
\begin{center}
    \centering
    \includegraphics[width=0.42\textwidth]{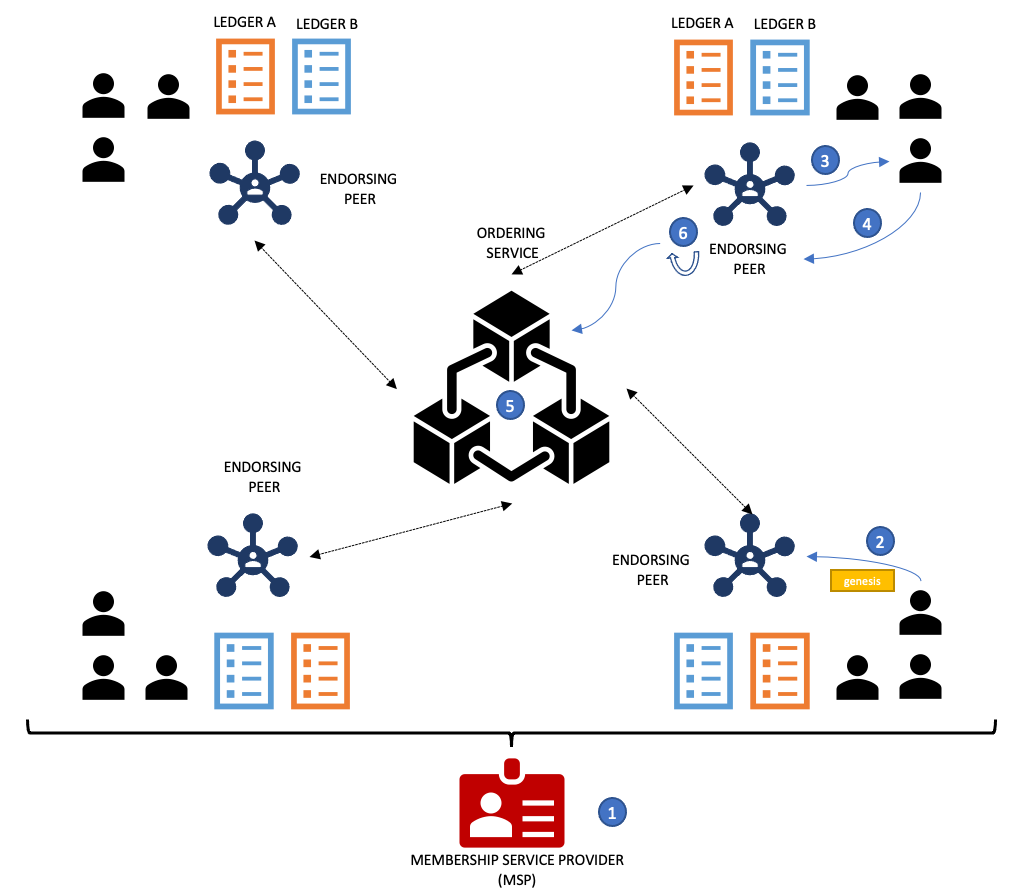}
    \caption{\scriptsize Platform Architecture for \textsc{Beas} implemented using HyperLedger Fabric Blockchain~\cite{androulaki,fabric}. Steps annotated as detailed in Scheme~\ref{alg:beasframe}.}
    \label{fig:arch}
\end{center}
\end{figure}

In traditional FL, each client $\mathcal{C}_{i}$ trains a \textit{local} model $\mathcal{M_L}^{i}$ with their own private data, and shares only the generated gradients with a centralized \textit{aggregator} that is responsible for storage and exchange of model gradients, and periodically merges previous \textit{local} model gradients to generate the new \textit{global} model $\mathcal{M_{G}}$. This is a vulnerable single point of failure as malicious aggregators can bias contributions of preferred clients over others~\cite{li} to skew the shared \textit{global} model. In a decentralized setting such as ours, we replace the role of the centralized aggregator with blockchain smart contracts~\cite{androulaki} for storage, exchange, and merging of model gradients. The framework architecture is depicted in Figure~\ref{fig:arch} and is implemented using the Fabric Blockchain~\cite{androulaki,fabric}. 

The framework is designed to be scalable to incorporate large numbers of participating clients, as well as allow parallel FL models on the same platform with minimum overheads. In our experiments, we assume that the participating nodes are organized in a tree-network topology with every endorsing peer (EP) connected to multiple clients for communication efficiency. However, our framework is completely decentralized and network topology agnostic; it does not necessitate any topological node hierarchy, and each EP can be set up to work with a single node if required.  The learning models are built using general-purpose ML APIs; hence \textsc{Beas} can support any model architecture that can be optimized using SGD. 

\begin{figure}[]
\begin{center}
    \centering
    \includegraphics[width=0.3\textwidth]{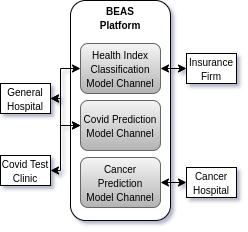}
    \caption{\textsc{Beas} uses multi-channel blockchain to enable parallel FL task execution.}
    \label{fig:multi-channel}
\end{center}
\end{figure}

Clients generate cryptographically anonymous identities using the HyperLedger Fabric Membership Service provider (MSP)~\cite{fabric}.  Multi-channel blockchain design is used to facilitate collaboration on multiple tasks simultaneously.  Every channel is set up with its own ML task, and a separate ledger. The ledger comprises a blockchain that stores model gradients as immutable and sequenced records in blocks, and a world-state database that stores the current state~\cite{androulaki,fabric}. We refer the reader to the Appendix for figures depicting the multi-channel blockchain design, as well as the block designs. 

\begin{figure}[]
\begin{center}
    \centering
    \includegraphics[width=0.35\textwidth]{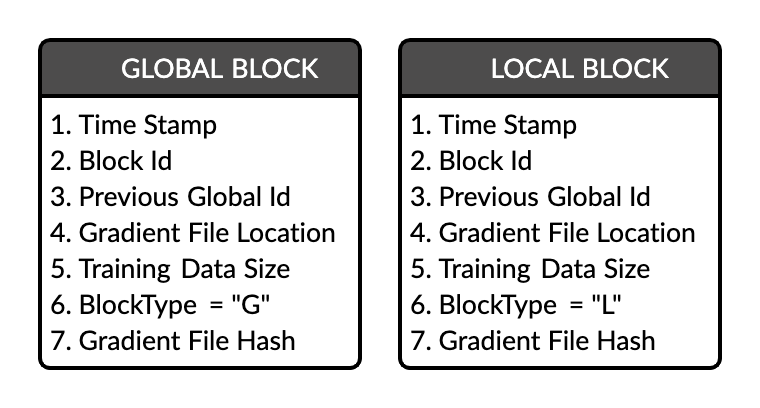}
    \caption{\textsc{Beas} blockchain block design.}
    \label{fig:blocks}
\end{center}
\end{figure}

To begin the training process, any client can set up a new channel, define the training parameters and the model architecture, and generate a genesis block by training on their own private data locally. The generated genesis block is appended to the channel's ledger as the first global block. Now, other clients can connect to this channel to use the global model as per their own requirements, or collaborate to improve the existing model with FL. In subsequent training rounds, clients request the Endorsing Peers (EP) that store a real-time copy of the ledger for the previous \textit{global} block on the ledger, and use this as a pre-training model. They train using their own private data to generate updated gradients, and then send them back to the EP to create a new \textit{local} block by executing smart contracts in an isolated environment. All the generated \textit{local} blocks at EPs are then sent to an Ordering Service, which establishes consensus on the order of the created blocks before sending them back to EP for commitment to the ledger. 

When a threshold $t$ of \textit{local} blocks is achieved on the ledger, a gradient merging smart contract is triggered to merge all the received \textit{local} blocks using federated gradient averaging~\cite{mcmahan} to generate a new \textit{global} block. Since different clients have varying sizes of datasets that are not independent and identically distributed (Non-i.i.d), the merge contract weighs the contribution of each \textit{local} block upon the size of its training dataset. 

We also limit the amount of data that can be used by any client to train in a given \textit{local} round to small chunks, and restrict \textit{local} training to a low number of epochs. That is, clients divide their datasets into small subsets that are used for training in subsequent rounds, instead of training on the entire dataset in a single round of \textit{local} training. This is done to ensure that the \textit{local} models do not pre-maturely converge while training, as that can lead to forgetting of features learned in previous rounds. 

% Constant collaborative contributions of trained models from participating clients is essential to improve the shared model performance. This necessitates a fair reward mechanism to motivate users for sharing local updates. \textsc{Beas} can be designed to include a token-based reward scheme to motivate users for collaboration, where users are given rewards for their contributions but have to pay a small fee for every access of the global model API as per their ML task requirements. Moreover, by setting up new channels with their own ML models, client can earn a small percentage from all transaction fees generated on that channel. Here, the fees can be based upon one or more of the following factors: (1) Performance of local model, (2) Amount of computational and network resources used for training the local model, (3) Continued honest behavior of the client. 
% \begin{inparaenum}[(1)]
%     \item Performance of local model.
%     \item Amount of computational and network resources used for training the local model.
%     \item Continued honest behavior of the client.
% \end{inparaenum}

\subsection{Adversarial Poisoning}

Malicious clients can tamper with their \textit{local} training datasets to perform data poisoning attacks~\cite{jagielski,bagdasaryan} that can introduce backdoors into the shared \textit{global} model. In fact, a single poisoned gradient can deviate the merged gradient significantly. Centralized FL frameworks majorly rely on server-owned validation datasets to evaluate \textit{local} gradient performance before aggregating them into the \textit{global} model, rejecting updates that perform poorly~\cite{xie}. However, in a privacy-sensitive decentralized setting such as ours, detecting anomalous updates without using a publicly available testing dataset is required.

% \begin{figure}[ht]
% \begin{center}
%     \centering
%     \includegraphics[width=5cm]{figure/adv.png}
%     \caption{Impact of poisoned (or faulty) gradients on model aggregation~\cite{xie}}
%     \label{fig:adv}
% \end{center}
% \end{figure}

We use the \textit{FoolsGold}~\cite{fg} (FG) defense protocol to address data poisoning attacks. In FL, training data is assumed to be non-IID. FG relies on similarity between client updates to distinguish honest participants from act-alike sybils by comparing the similarity of their gradient updates: sybils will contribute updates that are closer to each other than those among honest clients. It adapts the learning rate per client based on the similarity in indicative features amongst clients in any given round. As the framework faces larger groups of adversaries, it has more information to more reliably detect similarity between sybils. Unlike other defenses, FG does not require knowledge of the number of sybil attackers, and does not require modifications to the client-side protocol.

Unfortunately, as FG works by detecting similarity in gradients amongst Sybils, it fails to detect solitary attackers. For this case, \textsc{Beas} complements FG with Multi-KRUM (MK)~\cite{blanchard}. Multi-KRUM~\cite{blanchard} is a byzantine-resilient gradient aggregation algorithm that can address data poisoning attacks. In every federated round, it scores each \textit{local} block based on its deviation from every other submitted blocks. It guarantees resiliency from $f$ malicious updates out of $n$ total updates, when $2f+2 <n$. For update $V_i \ \forall\  i\in [1, n]$, $Score(V_i)$ is calculated as the sum of euclidean distances between $V_i$ and $V_j$, where $V_j$ denotes the $n-f-2$ closest vectors to $V_i$ as: $Score(V_i) = \sum_{i\rightarrow j} \mid\mid V_i - V_j \mid\mid^2$. Here, $i \rightarrow j$ denotes the fact that $V_j$ belongs to $n-f-2$ closest vectors to $V_i$. The $n-f$ updates with the lowest scores are selected for aggregation, and the rest are discarded. We refer the reader to~\cite{blanchard} for security guarantees and convergence analysis of Multi-KRUM, and to~\cite{fg} for detailed description of the FoolsGold protocol.

To evade anomaly detection algorithms, malicious clients can also manipulate the training algorithm, their model parameters, or directly the model gradients, to generate poisoned model gradients. Bagdasaryan et al.~\cite{bagdasaryan} demonstrates a backdoor model poisoning attack that uses \textit{constrain-and-scale} techniques to generate gradients that do not look anomalous, but can introduce backdoors in the \textit{global} model after merging.  The generated model can achieve high accuracy on both -- the main task, and an attacker-chosen backdoor subtask. Bagdasaryan et al.~\cite{bagdasaryan} demonstrates that DP based techniques with low clipping bounds and high noise variance can render the backdoor attack ineffective, but significantly decrease the accuracy of the \textit{global} model on its main task. To minimize the efficacy of model poisoning attacks with DP, but without impacting model performance, we experiment with gradient pruning where gradients with small magnitudes are pruned to zero~\cite{tsuzuku,lin}. Our experimental results show that as more and more gradient values are pruned, the sparsity in the model gradients increases: this has minimal impact on the main task, but minimizes the accuracy on the backdoor subtask under a model poisoning attack~\cite{blanchard} due to gradient obfuscation, as it makes \textit{constrain-and-scale} hard. We conclude that gradient pruning is an effective defense against model poisoning attacks on the backdoor subtask, and include pruning in the \textit{local} training rounds for \textsc{Beas}.

\subsection{Privacy Leakage}

In FL, client data is not directly shared with a centralized aggregator; to prevent leakage of data privacy, only the gradients are shared. However, ~\cite{song} demonstrates that a malicious aggregator can reveal sensitive information even from shared gradients that inadvertently retain features from the data used to train them. ~\cite{zhu} further proposes a Deep Leakage from Gradient (DLG) attack where an adversary can reconstruct the training dataset and labels from gradients by first generating `dummy' inputs and labels, and then performing the usual forward and backward pass using them to generate dummy gradients. Now, instead of optimizing model weights as in a typical learning setting, the adversary optimizes the dummy inputs and labels by minimizing the deviation between the dummy and real gradients, with the goal to make the dummy data close to the original ones, thereby revealing private information~\cite{zhu}.

Prior work proposes the use of cryptographic techniques such as secure multi-party computation (MPC)~\cite{perry2014systematizing}, Homomorphic Encryption~\cite{rivest1978}, Functional Encryption~\cite{xu}, and Secret Sharing~\cite{shamir} for the secure exchange of model gradients. However, cryptographic techniques involve significant computational overheads, and perform poorly in scale~\cite{phong}. Moreover, since cryptographic techniques encrypt shared gradients to prevent leakage of privacy, the gradients cannot be evaluated using anomaly detection algorithms. This renders the framework vulnerable to poisoning attacks. Similarly, DP techniques such as the addition of DP-noise and clipping of gradient values to limit privacy leakage by obfuscating model gradients have been explored in prior work, but impact accuracy~\cite{lyu,abadi,dwork}. 

Our experimental results demonstrate that gradient pruning is more effective than other DP-based techniques to preserve user data privacy under FL while ensuring minimal impact on model performance, but has not been used in recent work. To evaluate the extent of possible leakage of privacy, we implement the DLG Attack ~\cite{zhu} with three different techniques -- the addition of noise, gradient value clipping, and gradient pruning. We observe that gradient pruning helps minimize reconstruction of sensitive information of the training dataset from shared gradients. Moreover, pruning has minimal impact on model performance compared to existing noise and clipping based DP techniques.
\section{Experimental Evaluation}\label{sec:experiment}

In this section, we experimentally evaluate \textsc{Beas}’s performance and present our empirical results, followed by a security and privacy analysis of the framework. 

\subsection{Implementation Details}

We implement the \textsc{Beas} framework using the HyperLedger Fabric \textit{v2.2} Blockchain \cite{androulaki,fabric}, with the smart-contracts and client applications programmed using Node.js. All the experiments are performed using a Linux server with Intel Xeon E3-1200 v5 CPUs and GeForce RTX 2070 GPU, with 32 GB RAM. All client nodes are executed in isolated docker containers, and are connected using a 2Gbps bandwidth virtual network with an average network delay of 0.2ms. 

\subsection{Experimental Setup}

Unless otherwise stated, we set up our test blockchain network with 5 endorsing peers (EP), each of which is connected to participating clients in two settings: (a) 4 clients per EP ($N$ = 4 $\times$ 5 = 20 Clients), and (b) 10 clients per EP ($N$ = 10 $\times$ 5 = 50 Clients), as our default network setting. In every training round, clients train a \textit{local} model using a subset of their data (of cluster size~$\textit{c})$, and send the generated gradients to the EP as \textit{local} blocks. When a threshold $t$ of \textit{local} blocks is achieved on the ledger, the merge chaincode is triggered to generate the new \textit{global} block by aggregating the previous \textit{local} blocks using federated averaging~\cite{mcmahan}. The merging threshold parameter $t$ is kept low to ensure that the number of global training rounds increases, which has better impact on model performance than merging more gradient contributions per round as per our experiments. The value of cluster size $c$ is set in a way that ensures at least 50 global training rounds can be achieved with the resultant dataset split. The cluster size $c$ and the threshold $t$ are defined in the following subsections based on the dataset used. 

\subsection{Datasets and Model Configurations}

We report our evaluation of the \textsc{Beas} framework using 3 popular and standard datasets with standard model configuration as follows:
% : (a) MNIST handwritten digits dataset~\cite{lecun-mnisthandwrittendigit-2010}, (b) Malaria Cell Image dataset~\cite{malaria}, (c) CIFAR-10 object recognition dataset~\cite{cifar-10}, and (d) Wisconsin Breast Cancer dataset~\cite{wisconsin}.

\subsubsection{MNIST~\cite{lecun-mnisthandwrittendigit-2010}.} \label{subsubsec:mnist}
% \noindent
% \textbf{MNIST~\cite{lecun-mnisthandwrittendigit-2010}.}
Consists of 60,000 training images resized to (28, 28, 1), along with 10,000 testing images. Model used is a feed-forward CNN model. Following~\cite{mcmahan,bagdasaryan}, we randomly split the dataset amongst the clients with an unbalanced sample from each class using a Dirichlet distribution~\cite{minka} with hyperparameter 0.9 to simulate \textit{non-i.i.d.} distribution. Merging threshold $t$ is set as 5, and \textit{local} training is done for 5 epochs with the learning rate of 0.1 using a randomly selected cluster of size $c = 250$ from the clients data subset, with standard batch size 32. 

\subsubsection{Malaria Cell Image~\cite{malaria}.}
% \noindent
% \textbf{Malaria Cell Image~\cite{malaria}.}
Consists of 27,558 images, resized to (50, 50, 3). We set aside the 20\% of the dataset for testing. Model used is a feed-forward CNN model. We split the dataset as done for MNIST above. Merging threshold $t$ is set as 5, and \textit{local} training is done for 3 epochs with the learning rate of 0.1 using a randomly selected cluster of size $c = 100$ from the clients data subset, with standard batch size 32. 

\subsubsection{CIFAR-10~\cite{cifar-10}.}
% \noindent
% \textbf{CIFAR-10~\cite{cifar-10}.}
Consists of 50,000 images, resized to (32, 32, 3), along with 10,000 testing images. Model used is a feed-forward CNN model. To obtain the accuracy results, we execute a simulated \textsc{Beas} (using Tensorflow) with approximated activation functions and a fixed-precision. We split the dataset as done for MNIST above. Merging threshold $t$ is set as 5, and \textit{local} training is done for 5 epochs with learning rate of 0.1 using a randomly selected cluster of size $c = 150$ from the clients data subset, with standard batch size 32.

% \subsubsection{Wisconsin Breast Cancer ~\cite{wisconsin}}
% \textbf{Wisconsin Breast Cancer~\cite{wisconsin}.}
% The dataset consists of 569 data points with 32 attributes for our classification task. The training model is a feed-forward neural network. We split the dataset as done for MNIST dataset above. Finally, we set the merging threshold $t$ as 3, and \textit{local} training is done for 2 epochs with the learning rate of 0.1 using a randomly selected cluster of size $c = 10$ from the clients data subset, with batch size 5, as the available data per client reduces after splitting amongst the participants. %Moreover, the number of epochs is set low to ensure that the \textit{local} models do not pre-maturely converge while training, as that can lead to forgetting of features learned in previous rounds. 

\subsection{Experimental Results}

We evaluate \textsc{Beas} in terms of (a) the accuracy of the resultant \textit{global} model, and (b) the scalability of the framework. Tables~\ref{table:beas-exp} show the results obtained in our two settings with $N=20$ and $N=50$ respectively. The accuracy column compares the accuracy obtained using \textsc{Beas} with a centralized ML approach, where the training dataset is available in a centralized setting. The execution time per local round, as well as the overall execution time are also mentioned. 
% We refer the reader to the Appendix for the accuracy graphs.

\begin{figure*}[ht]
\centering
\includegraphics[width=1\textwidth]{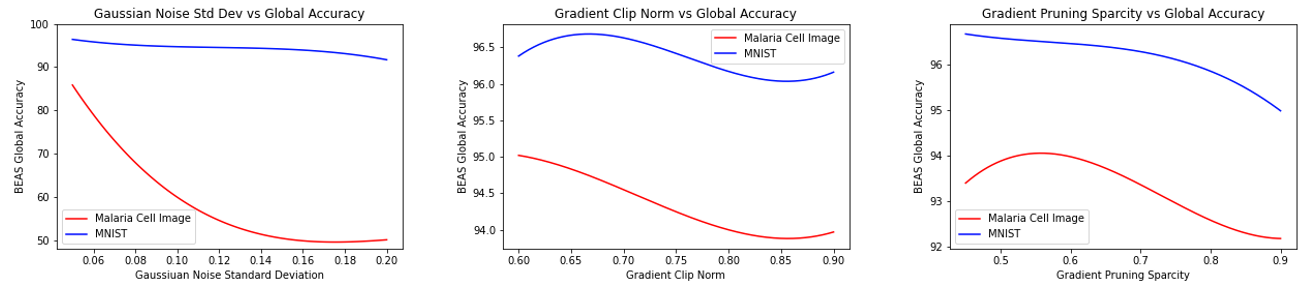}
\caption{Impact of differential privacy parameter settings on \textsc{Beas} global model accuracy; Dataset: Malaria Cell Image~\cite{malaria}, and MNIST~\cite{lecun-mnisthandwrittendigit-2010}.}
\label{fig:dp-graphs}
\end{figure*}

\begin{figure*}[ht]
\centering
\includegraphics[width=1\textwidth]{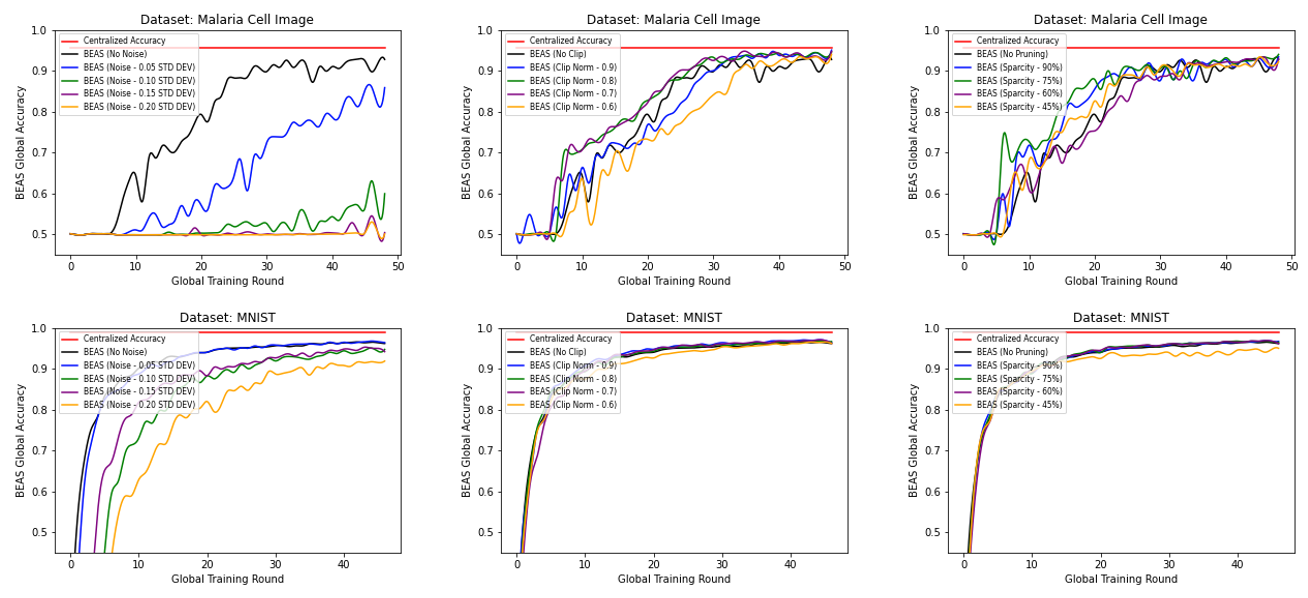}
\caption{Impact of different differential privacy techniques on \textsc{Beas} global model accuracy; Dataset: Malaria Cell Image~\cite{malaria}, and MNIST~\cite{lecun-mnisthandwrittendigit-2010}.}
\label{fig:dp-graphs-comp}
\end{figure*}

\begin{table*}[]
% \scriptsize
\centering
\caption{\textsc{Beas}’s accuracy and execution times for $N = 20$ and $N$ = 50 clients.}
\label{table:beas-exp}
\begin{tabular}{|l|c|c|c|c|c|c|c|}
\hline
\multirow{3}{*}{Dataset} & \multirow{3}{*}{\shortstack{Centralized \\ Accuracy \\ (\%)}} & \multicolumn{3}{c|}{$N = 20$}                                              & \multicolumn{3}{c|}{$N = 50$}                                              \\ \cline{3-8} 
                         &                                       & \multirow{2}{*}{\shortstack{Accuracy \\ (\%)}} & \multicolumn{2}{l|}{Avg. Execution Time (s)} & \multirow{2}{*}{\shortstack{Accuracy \\ (\%)}} & \multicolumn{2}{l|}{Avg. Execution Time (s)} \\ \cline{4-5} \cline{7-8} 
                         &                                       &                           & Local Training           & Overall           &                           & Local Training           & Overall           \\ \hline

MNIST & 95.53  & 92.74  & 1.89 & 524  & 90.11  & 1.89  & 726   \\ \hline

Malaria & 98.89  & 96.16 & 2.18  & 967 & 92.81  & 2.18   & 1276   \\ \hline

CIFAR-10 & 72.81  & 61.03 & 38  & 21608  & 63.76  & 38  & 25966   \\ \hline

% WBC~\cite{wisconsin} & 97.33  & 97.29  & 0.07  & 76   & -  & -  & -  \\ \hline
\end{tabular}
\end{table*}

\section{Security and Privacy Analysis}\label{sec:secpriana}

We analyze the security and privacy of our proposed framework from (a) reconstruction attacks that are threat to client data privacy, and (b) two types of adversarial attacks -- data poisoning attack~\cite{jagielski} and model poisoning attack~\cite{bagdasaryan}. %Table~\ref{table:companalysis} in \S\ref{sec:companalysis} summarizes different approaches with respect to recent state-of-the-art solutions. We refer the reader to \S\ref{sec:expfig} for the accuracy graphs.

\subsection{Privacy Guarantees}\label{subsec:privacy} 

% Zhu et al.~\cite{zhu} demonstrates that it is possible to reconstruct sensitive information from generated model gradients. As \textsc{Beas} relies on sharing of model gradients instead of real data, we experiment with different differential privacy techniques that can be used to ensure that the presence (or absence) of any given element in a dataset does not result in the generation of vastly different model gradients (which could potentially be used to reveal private information).

We experiment with the Deep Leakage from Gradient Attack (DLG Attack) proposed by~\cite{zhu} to understand the extent of possible leakage from shared gradients. We also examine the impact of Gaussian noise addition, gradient value clipping, and gradient pruning to minimize this leakage. To simulate a DLG Attack~\cite{zhu}, we begin by training an ML model to generate the gradients $G_{real}$ using our dataset $D_{real}$. Then, we generate a pair of dummy inputs and labels  $D_{dummy}$ using random noise, and train the same ML model to get dummy gradients $G_{dummy}$. Now, instead of optimizing $G_{dummy}$, we optimize $D_{dummy}$ using a loss function $\mid \mid D_{real} - D_{dummy}\mid \mid^2$ that minimizes the distance between the real gradients and the dummy gradients. Matching the gradients makes the dummy data close to the original ones. Our experiments show that as the optimization completes, the private training data can be completely revealed (We refer the reader to the Appendix for the reconstruction experiment figures). 

We experiment with the following standard DP based schemes \cite{zhu} to reduce adversarial reconstruction ability of sensitive information from shared model gradients on \textsc{Beas}: (a) {Gaussian Noise Addition:} The mechanism adds appropriately chosen random noise to the generated gradients to produce obfuscated gradients. However, our experiments show that although it successfully prevents reconstruction at high standard deviation values, it negatively impacts the model performance. (b) {Gradient Value Clipping:} This forces element-wise gradient values to a specific minimum (or maximum) value if the gradient exceeded an expected range~\cite{brownlee}. However, our experiments show that it does not minimize leakage of private data from shared model gradients as the ground truth image can easily be reconstructed. (c) {Gradient Pruning:} Involves pruning of gradients with small magnitudes to zero to increase the sparsity between gradients values. In effect, gradient values with insignificant magnitudes are pruned, but gradient values with higher magnitude contributions are preserved. Our experiments show that as we increase gradient sparsity to more than 50\%, the recovered images in a DLG attack~\cite{zhu} are no longer recognizable, successfully preventing leakage of sensitive information from shared gradients. 

% \textit{Addition of Noise:} When the query is a function $f$, and the database is $X$, the true answer is the value $f(X)$. The mechanism $K$ adds appropriately chosen random noise to the true answer to produce what we call the response~\cite{dwork}. Generally, a Gaussian or Laplace distribution is used for generating the noise for gradient perturbation.

% % \noindent
% \textit{Gradient Clipping:} Gradient value clipping involves clipping the derivatives of the loss function to have a given value if a gradient value is less than a negative threshold, or more than the positive threshold~\cite{brownlee}.

% % \noindent
% \textit{Gradient Pruning:}  In gradient pruning, gradients with small magnitudes are pruned to zero. The sparsity in the gradients increases as more and more gradient values are pruned, making attacks on data privacy hard~\cite{tf,tsuzuku,lin}. 

\begin{table}[!ht]
\centering
\scriptsize
\caption{\label{table-dp-pars} Impact of different differential privacy parameter settings on \textsc{Beas} global model accuracy}
\begin{tabular}{|c|c|c|c|}
\hline
\multicolumn{1}{|c|}{\multirow{2}{*}{Framework}} & \multicolumn{1}{c|}{\multirow{2}{*}{Parameters}} & \multicolumn{2}{c|}{Accuracy (\%)} \\ \cline{3-4} 
\multicolumn{1}{|c|}{}                           & \multicolumn{1}{c|}{}                            & MNIST        & Malaria        \\ \hline

Centralized                                                                                                      & -              & 95.53             & 98.89          \\ \hline
BEAS                                                                                                             & -              & 92.74             & 96.16          \\ \hline
\multirow{4}{*}{\begin{tabular}[c]{@{}c@{}}BEAS + Gaussian \\ Noise Addition (Std Dev)\end{tabular}} 
                                                                                                                 & Noise (0.10)   & 59.92             & 94.7           \\ \cline{2-4} 
                                                                                                                 & Noise (0.15)   & 50.41             & 94.16          \\ \cline{2-4} 
                                                                                                                 & Noise (0.20)   & 50.07             & 91.68          \\ \hline
\multirow{4}{*}{\begin{tabular}[c]{@{}c@{}}BEAS + Gradient \\ Clipping (Clip Norm)\end{tabular}}        & Clip (0.90)    & 95.02             & 96.38          \\ \cline{2-4} 
                                                                                                                 & Clip (0.80)    & 94.55             & 96.63          \\ \cline{2-4} 
                                                                                                                 & Clip (0.70)    & 94.7              & 96.17          \\ \hline
\multirow{4}{*}{\begin{tabular}[c]{@{}c@{}}BEAS + Gradient \\ Pruning (Sparcity)\end{tabular}}          & Pruning (0.90) & 93.39             & 96.67          \\ \cline{2-4} 
                                                                                                                 & Pruning (0.75) & 93.97             & 96.46          \\ \cline{2-4} 
                                                                                                                 & Pruning (0.60) & 92.95             & 96.11           \\ \hline
\end{tabular}
\end{table}

\begin{table}[!ht]
\centering
% \scriptsize
\caption{\label{table:mk} \textsc{Beas} accuracy with FoolsGold (FG)~\cite{fg} and Multi-KRUM (MK)~\cite{blanchard} under Label Flipping attack~\cite{labelflip} for different number of adversaries and $(N=20)$; Dataset: Malaria Cell Image~\cite{malaria}.}
% \begin{tabular}{|c|c|c|c|c|}
% \hline
%                         & \multicolumn{4}{c|}{\bf Number of Adversaries} \\ \hline
% \textbf{\% Blocks  Rejected (m)} & 0         & 1        & 5        & 10       \\ \hline

\begin{tabular}{|c|c|c|c|c|}
\hline
\multirow{2}{*}{Defense} & \multicolumn{4}{l|}{Number of Adversaries} \\ \cline{2-5} 
                                        & 0        & 1        & 5        & 10        \\ \hline

NIL                   & 96.16     & 96.02    & 82.88    & 57.20    \\ \hline
MK                    & 94.22     & 94.60    & 91.17    & 72.11    \\ \hline
FG                    & 95.63     & 82.11    & 87.50    & 85.72    \\ \hline
MK + FG               & 94.16     & 90.26    & 87.24    & 83.66    \\ \hline
\end{tabular}
\end{table}

Table~\ref{table-dp-pars} shows the impact of different parameter settings for these DP techniques on the global model accuracy for \textsc{Beas}. We refer the reader to the Appendix for the accuracy graphs. We observe that gradient pruning has minimal impact on model performance compared to other DP based techniques, but successfully minimizes reconstruction of training inputs from shared model gradients under the DLG Attack. This is consistent with the results obtained by~\cite{zhu}, where the impact of DP techniques on privacy leakage is evaluated. We refer the reader to the Appendix for the reconstruction experiment figures.

\begin{figure}[!ht]
\begin{center}
    \centering
    \includegraphics[width=0.47\textwidth]{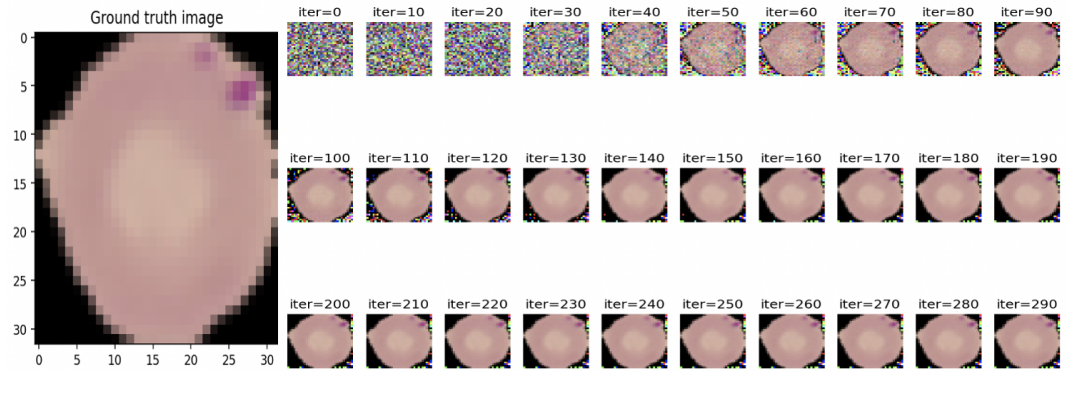}
    \caption{Private data reconstruction using
    Deep Leakage from Gradient Attack~\cite{zhu}; as the optimization completes, $D_{dummy}$ reveals the ground truth training image. 
    Dataset: Malaria Cell Image~\cite{malaria}}
    \label{fig:dlg-main}
\end{center}
\end{figure}

\begin{figure}[!ht]
    \centering
    \includegraphics[width=0.47\textwidth]{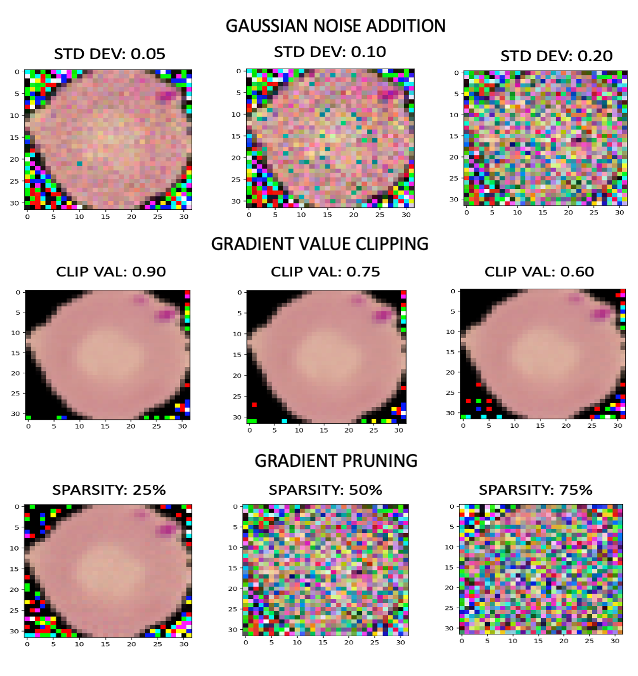}
    \caption{Differential Privacy Techniques against Deep Leakage from Gradient Attack~\cite{zhu} to reconstruct training data from shared model gradients; Reconstruction results after 300 iterations.}
    \label{fig:dlg-rec}
\end{figure}

\subsection{Adversarial Security}\label{subsec:security} 

Malicious users can train their local models using manipulated ``poisoned'' datasets to induce a specific outcome in the ML task at inference time~\cite{jagielski}. To simulate a data poisoning, we execute a label flipping attack~\cite{nguyen,tolpegin,labelflip} on the Malaria Cell Image dataset~\cite{malaria}. Label flipping attacks are a special type of data poisoning attacks, wherein the attacker can manipulate the class labels assigned to a fraction of training inputs. Label flipping attacks can significantly diminish the performance of the system, even if the attacker’s capabilities are otherwise limited~\cite{labelflip}. We experiment with 1, 5, and 10 adversarial participants in our attack setup, who modify their training datasets by flipping the class label from parasitized to uninfected, and vice versa. The aim is to make the \textit{global} model more likely to classify the label in the learning task incorrectly~\cite{nguyen,tolpegin}. To detect and reject malicious updates, we implement FoolsGold\cite{fg} and Multi-KRUM~\cite{blanchard}. Table~\ref{table:mk} shows the impact of label flipping attack with different number of adversarial participants. (Note that with Multi-KRUM, the accuracy decreases even without any adversarial clients. This is because the number of training samples also decreases with high rejection of blocks.)

% \begin{table}[ht]
% \centering
% \scriptsize
% \caption{\scriptsize \label{table:mk} \textsc{Beas} accuracy with Multi-KRUM~\cite{blanchard} under Label Flipping attack~\cite{labelflip} for different number of adversaries and $(N=20)$; Dataset: Malaria Cell Image~\cite{malaria}.}
% % \begin{tabular}{|c|c|c|c|c|}
% % \hline
% %                         & \multicolumn{4}{c|}{\bf Number of Adversaries} \\ \hline
% % \textbf{\% Blocks  Rejected (m)} & 0         & 1        & 5        & 10       \\ \hline

% \begin{tabular}{|c|c|c|c|c|}
% \hline
% \multirow{2}{*}{\% Blocks Rejected (m)} & \multicolumn{4}{l|}{Number of Adversaries} \\ \cline{2-5} 
%                                         & 0        & 1        & 5        & 10        \\ \hline

% 0\% (No Multi-KRUM~\cite{blanchard})                   & 96.16     & 96.02    & 82.88    & 57.20    \\ \hline
% 25\%                    & 94.22     & 94.60    & 91.17    & 72.11    \\ \hline
% 50\%                    & 79.21     & 74.38    & 70.10    & 76.06    \\ \hline
% \end{tabular}
% \end{table}

\begin{table}[!ht]
\centering
\scriptsize
\caption{\label{table:pixel} \textsc{Beas} accuracy on main task and backdoor subtask with different differential privacy techniques under Pixel Pattern Backdoor Model Poisoning attack~\cite{bagdasaryan} for different number of adversaries and $(N=20)$; Dataset: Malaria Cell Image~\cite{malaria}.}
\begin{tabular}{|c|c|c|c|c|c|c|}
\hline
Framework & \multicolumn{3}{c|}{\begin{tabular}[c]{@{}c@{}}Main Task \\ Accuracy (\%)\end{tabular}} & \multicolumn{3}{c|}{\begin{tabular}[c]{@{}c@{}}Backdoor Task \\ Accuracy (\%)\end{tabular}} \\ \hline
                                                                           
\begin{tabular}[c]{@{}c@{}}Adversaries\\ per Round\end{tabular}            & 0                           & 1                           & 5                           & 0                             & 1                            & 5                            \\ \hline
BEAS                                                                       & 96.16                       & 95.81                       & 96.08                       & 11.06                         & 28.20                        & 61.85                        \\ \hline
\begin{tabular}[c]{@{}c@{}}BEAS + \\ Noise (0.05)\end{tabular}    & 85.84                       & 84.66                       & 82.10                       & 09.76                         & 19.44                        & 49.16                        \\ \hline
\begin{tabular}[c]{@{}c@{}}BEAS + \\ Clipping (0.80)\end{tabular} & 94.55                       & 94.16                       & 93.60                       & 11.33                         & 27.21                        & 62.70                        \\ \hline
\begin{tabular}[c]{@{}c@{}}BEAS + \\ Pruning (0.60)\end{tabular}  & 92.95                       & 92.67                       & 92.88                       & 10.20                         & 22.46                        & 43.88                        \\ \hline
\end{tabular}
\end{table}

To evade anomaly detection algorithms like Multi-KRUM~\cite{blanchard}, malicious clients can introduce backdoors into the joint model by manipulating the training algorithms, the model parameters, or the model gradients, to generate poisoned model gradients. ~\cite{bagdasaryan} proposes a backdoor model poisoning attack that uses \textit{constrain-and-scale} techniques to generate poisoned model gradients that do not look anomalous, and replace the \textit{global} model after aggregating with the other participants’ models.  The generated model can achieve high accuracy on both the main task, and an attacker-chosen backdoor subtask. However, the author also shows that DP based techniques with low clipping bounds and high noise variance can render the backdoor attack ineffective. However, our experimental results show that existing DP techniques such as addition of noise and clipping the gradient norms significantly impact the accuracy of the \textit{global} model on its main task. 

Since \textsc{Beas} uses gradient pruning based DP, we compare its efficacy in minimizing the impact of a model poisoning attack w.r.t other differential privacy techniques. To simulate a model poisoning attack, we implement a pixel pattern model poisoning attack~\cite{bagdasaryan} using the Malaria Cell Image Dataset~\cite{malaria}. Here, the attacker modifies a subset of the training image's pixels in order for the model to incorrectly classify the modified image on a backdoor task, and then uses \textit{constrain-and-scale} techniques to evade gradient anomaly detection algorithms such as Multi-KRUM~\cite{blanchard}. 

Table~\ref{table:pixel} shows the performance of the generated \textit{global} model on the main task for 0, 1, and 5 adversaries per global iteration, along with the performance of the generated model on the backdoor subtask. We see that increasing the number of adversaries has minimal impact on the accuracy on the main task, but increases the accuracy of the backdoor subtask. We also note that Gradient Pruning minimizes the accuracy on the backdoor subtask, with minimal impact on the performance of the global model. We conclude that gradient pruning is the most effective defense against model poisoning attacks on the backdoor subtask.

\section{Conclusion}\label{sec:conclusion}

From adversarial poisoning attacks to client data privacy leakage, traditional FL setups face several challenges. We propose \textsc{Beas}, a blockchain based framework which overcomes these challenges, and conduct extensive experiments with multiple datasets and observe promising results. \textsc{Beas} achieves approximately 92.72\% accuracy on the MNIST~\cite{lecun-mnisthandwrittendigit-2010} dataset, and 96.16\% accuracy on the Malaria Cell Image~\cite{malaria} dataset while training on DNNs. We also scale \textsc{Beas} with a large number of participants (20 to 50), where it achieves a reasonable accuracy of 90.11\% on MNIST. To ensure detection and rejection of malicious nodes that try to poison the global model with tampered datasets, we implement anomalous gradient detection~\cite{blanchard,fg}. We also implement gradient pruning to maximize privacy protection without sacrificing model utility loss, which is the case when using traditional DP techniques. Gradient pruning also helps us reduce the efficacy of model poisoning attacks~\cite{bagdasaryan}.  

In the future, we wish to conduct tests using synthetic data for effective privacy preservation, and ablation studies where we observe how removal (or addition) of specific features in the model affect performance in a decentralized FL setting. 
% Moreover, since Multi-KRUM can be evaded by more advanced poisoning attacks proposed in recent work, we wish to experiment and implement more secure anomaly detection protocols. 
We also plan to deploy our framework via open-source channels for different academic and industrial purposes to observe its working real-time.

%
% ---- Bibliography ----
%
\bibliography{ppai-ref}

\appendix
\section{Appendix}\label{section:appendix}

\subsection{More Detail on Poisoning Attacks}\label{subsec:poiatk}

% \subsection{Poisoning Attacks}\label{subsec:poiatk}

In poisoning attacks, the adversary $\mathcal{A}^c$ manipulates the local models $W_i$ of $k<\frac{K}{2}$ clients to obtain poisoned models $W'_i$ which are then aggregated
into the global model $G_t$ and affect its behavior. Poisoning attacks can be divided into untargeted and targeted attacks. In untargeted attacks, the adversary’s goal is merely to impact (deteriorate) the benign performance of the aggregated model, while in targeted attacks (also called backdoor attacks), the
adversary wants the poisoned model $G'_t$ to behave normally on all inputs except for specific attacker-chosen inputs $x \in \mathcal{I}_{\mathcal{A}^c}$ for which attacker-chosen (incorrect) predictions should be output~\cite{nguyen}.

\paragraph{Data Poisoning Attack}\label{subsec:dpa}

In a data poisoning attack, the adversary $\mathcal{A}^c$ adds manipulated ”poisoned” data to the training data used to train local model $W_i$~\cite{nguyen}. For this, the adversary can implement a label flipping attack~\cite{tolpegin}: given a source class $c_{src}$ and a target class $c_{target}$ from $\mathcal{C}$, each malicious participant $\mathcal{A}_i$ modifies their dataset $D_i$ as follows: For all instances in $D_i$ whose class is $c_{src}$, change their class to $c_{target}$. We denote this attack by $c_{src} \rightarrow c_{target}$. The goal of the attack is to make the final global model $G_t$ more likely to misclassify the primary model task~\cite{nguyen,tolpegin}.

\begin{algorithm}[!ht]
\SetAlgoLined
\scriptsize
\textbf{Constrain-and-scale}($\mathcal{D}_{local}, {D}_{backdoor}$)

Initialize attacker’s model $X$ and loss function $l$

$X \leftarrow G^t$

$l \leftarrow \alpha \times \mathcal{L}_{class} + (1- \alpha) \times \mathcal{L}_{ano}$

\For{epoch $e \in E_{adv}$}{
    \If{$\mathcal{L}_{class}(X, {D}_{backdoor}) < \epsilon$}{
        \textit{break} \tcp*{\scriptsize Early stop, if model converges}
    } 
    \For{batch $b \in \mathcal{D}_{local}$}{
        $b \leftarrow$ replace($c, b, {D}_{backdoor}$)
        
        $X \leftarrow X - lr_{adv}\times \nabla l(X, b)$
    }
    \If{epoch e $\in$ step\_sched}{
        $lr_{adv} \leftarrow lr_{adv}/{step\_rate}$
    }
    $\widetilde{L}^{t+1} \leftarrow \gamma(X-G^t)+G^t$ \tcp*{\scriptsize Scale up the model before submission}
}
\Return{$\widetilde{L}^{t+1}$}

\caption{\scriptsize Constrain-and-scale~\cite{bagdasaryan} for model poisoning attack, where the objective function is modified by adding an anomaly detection term $\mathcal{L}_{ano}$; $\mathcal{L}_{model} = \alpha \times \mathcal{L}_{class} + (1-\alpha)\times \mathcal{L}_{ano}$. Hyperparameter $\alpha$ controls the importance of evading anomaly detection.}
\label{alg:cns}
\end{algorithm}

\paragraph{Model Poisoning Attack}\label{subsec:mpa}

In a model poisoning attack, the adversary $\mathcal{A}^c$ manipulates the training algorithms, their parameters, or directly manipulates (e.g., by scaling) the model $W'_i$. When performing the attack, the adversary seeks to maximize attack impact while ensuring the distance (e.g., Euclidean distance) between poisoned models $W'$ and benign models
$W$ remains below the detection threshold $\epsilon$ such as Multi-KRUM~\cite{blanchard} of the aggregator to evade possible anomaly detection performed by the aggregator on individual clients’ model updates~\cite{bagdasaryan,tolpegin,nguyen}.

Bagdasaryan et al.~\cite{bagdasaryan} propose a model poisoning attack that uses \textit{constrain-and-scale} techniques to create a model that does not look anomalous and replaces the global model after averaging with the other participants’ models (see Algorithm~\ref{alg:cns}).

\end{document}